# Photovoltaic Effect in Individual Asymmetrically Contacted Lead Sulfide Nanosheets

Sedat Dogan, Thomas Bielewicz, Vera Lebedeva, Christian Klinke*

*Institute of Physical Chemistry, University of Hamburg,*

*Grindelallee 117, 20146 Hamburg, Germany*

**Abstract**

*Solution-processable, two-dimensional semiconductors are promising optoelectronic materials which could find application in low-cost solar cells. Lead sulfide nanocrystals raised attention since the effective band gap can be adapted over a wide range by electronic confinement and observed multi-exciton generation promises higher efficiencies. We report on the influence of the contact metal work function on the properties of transistors based on individual two-dimensional lead sulfide nanosheets. Using palladium we observed mobilities of up to 31 cm²/Vs. Furthermore, we demonstrate that asymmetrically contacted nanosheets show photovoltaic effect and that the nanosheets' height has a decisive impact on the device performance. Nanosheets with a thickness 5.4 nm contacted with platinum and titanium show a power conversion efficiency of up to 0.94 % (EQE 75.70 %). The results underline the high hopes put on such materials.*

* Corresponding author: klinke@chemie.uni-hamburg.de

Colloidal inorganic semiconductor nanocrystals have emerged as attractive materials for the fabrication of novel electronic devices.[1-5] The constituents are nanometer-sized particles suspended in a solvent with the aid of stabilizing ligands. The geometry of the nanocrystals has a strong impact on their physical properties, e.g. their effective band gap is a matter of their diameter and shape.[6-8] The tremendous progress in solution processability and spectral tunability based on quantum confinement of nanocrystals offers a promising route for low-cost manufacturing of electronic devices. Thus, nanocrystal film preparation (by spin-coating, dip-coating, Langmuir-Blodgett, etc.) has been tackled to produce field-effect transistors,[1,2,3,4,5,9] photo detectors,[10,11] and solar cells.[12-16] However, it is a great challenge to achieve remarkable device performances since the charge transport is hampered by unfavorable band alignment, traps, and in particular by stabilizing, insulating ligands between the nanoparticles which represent tunnel barriers limiting the electrical transport. A variety of approaches have been undertaken to improve the conductivity of nanocrystal films. Among them are removal of the ligands by hydrazine treatment,[1] the usage of shorter, down to atomic ligands,[17-20] the replacement by inorganic ones,[9] with band alignment,[21] or carbonization of the ligands.[22] One approach which avoids tunnel barriers ab initio is to synthesize larger, two-dimensional structures, which are still solution processable and tunable in their band properties by their height, e.g. in materials such as CdSe[23,24] and SnSe.[25] Recently, we introduced a colloidal synthesis for lead sulfide nanosheets.[26] They exhibit lateral dimensions of a couple of micrometers and thicknesses of a few nanometers. They are continuous in the lateral dimensions, thus naturally without tunnel barriers in the plane. In a further work, we showed that they can be used as conductive channel in field-effect transistors.[4] The pristine PbS nanosheets contacted with gold electrodes showed p-type behavior with a more pronounced switching behavior at lower temperatures. Additionally, they showed remarkable photoconductivity. With our studies we could demonstrate that PbS nanosheets can be used as easy processable transistors which outperform transistors based on untreated nanocrystal films. Lead sulfide nanocrystals raised also attention since the effective band gap can be adapted over a wide range by electronic confinement[27] and observed multi-exciton generation promises higher solar cell efficiencies.[28-30]

To build photovoltaic cells with high efficiencies and low manufacturing costs there have been intensive research efforts over the recent decades. Fabricating dye-sensitized solar cells, [31-33] organic solar cells, [34,35] and solution-processed colloidal nanocrystal solar cells[35,36] are among those concepts. Latter emerge as hot candidates for next-generation solar cells since their spectral properties can be tuned by modification of the nanocrystal size to the optimum value given by a tradeoff between a large band gap resulting in a high open-circuit voltage and a small band gap which collects more of the solar spectrum resulting in a higher current. One of the first solar cells based on inorganic nanocrystals were developed by the Alivisatos group in 2002.[12] Later the same group published an all-inorganic solar cell. [38] PbS nanocrystals are considered for photovoltaic applications due to their excellent photosensitivity in the near-infrared spectrum.[39] Schottky barrier solar cells based on PbS nanocrystals have been demonstrated by various groups.[40-45] In a typical architecture the PbS nanocrystals are embedded as photoactive layer between indium tin oxide (ITO) and thermally evaporated metal electrodes. The PbS nanocrystals form an Ohmic contact with the ITO electrode and a Schottky contact with the metal.

Here, we report that the interface between metal contacts and individual PbS nanosheets plays a crucial role for the transistor performance. Depending on the metal work functions either Ohmic contacts or Schottky barriers can be formed. We found that PbS nanosheets furnished with metals with comparatively large work function like gold, palladium, and platinum form Ohmic contacts and low work function metals like titanium form Schottky barriers at the interface hampering the hole transport. Based on those findings, we demonstrate that PbS nanosheets contacted with two different metals with a large difference in work function show photovoltaic effect.

Figure 1 shows a TEM and AFM images of individually contacted nanosheets which possesses a quite smooth surface and a thickness of the inorganic part of 5.4 nm. By using the particle-in-a-box approach we calculate an effective band gap of 0.68 eV. Examples of transfer characteristics of nanosheets contacted with the metals titanium, gold, palladium, and platinum can be found in Figure 2. In case that the nanosheets are contacted with titanium leads with a low work function of $\phi(Ti) = 4.33$ eV we found a comparatively small conductivity of $\sigma = 11.96$ mS/m averaged over all measured devices. Although clearly being a p-type device only a shallow switching behavior was observed in the transfer characteristics with an average $I_{ON}/I_{OFF}$ ratio of 3.26 and a field-effect mobility µ of 0.33 cm²/Vs. The output characteristics display also a defined p-type behavior (see Figure S1). In contrast, using high work function metals like Au (ϕ = 5.10 eV), Pd (ϕ = 5.12 eV), and Pt (ϕ = 5.65 eV) we observe larger currents and much more pronounced switching behavior. The transfer characteristics show also p-type behavior for all gold and palladium contacted devices. Using gold as contact material we found σ = 68.06 mS/m, $I_{ON}/I_{OFF}$ ratio of 13.33, and a field-effect mobility of 8.35 cm²/Vs in average. For palladium we calculated σ = 67.07 mS/m, $I_{ON}/I_{OFF}$ ratio of 9.78, and a field-effect mobility of 7.72 cm²/Vs in average. A few devices exhibit mobilities of over 20 cm²/Vs and one showed even 31.80 cm²/Vs. We interpret these results as sign for the formation of Ohmic contacts. For the PbS nanosheet transistors contacted with platinum we found that the majority of the devices shows n-type character with a few p-type ones. We obtained σ = 99.0 mS/m, $I_{ON}/I_{OFF}$ ratio of 25.29, and a field-effect mobility of 2.57 cm²/Vs. Table 1 summarizes the averaged key values of the devices.

The devices contacted with platinum show the highest $I_{ON}/I_{OFF}$ ratio (25.29) and with titanium the lowest one (3.26). Using gold and palladium contacts we got the highest level of field-effect mobility (µ (Au) = 8.35 cm²/Vs, µ(Pd) = 7.72 cm²/Vs at $V_g$ = -5.0 V) and for titanium the lowest one (µ(Ti) = 0.33 cm²/Vs). Furthermore, we calculated for each metal contact pair the average conductivity which was found to be in average 0.01 S/m for Ti, 0.07 S/m for Pd and Au. For Pt we calculate an averaged conductivity of 0.105 S/m. This shows that the charge carrier transparency increases with work function.

The alignment of the energy levels at the interface between metal contacts and semiconducting material is essential for the switching behavior of the PbS nanosheet transistor. Figure 3 presents idealized band diagrams of the devices. We calculated the valence and conduction band levels by the quantum mechanical particle-in-a-box approach (effective mass approximation). These considerations confirm that after Fermi level equilibration Ohmic contacts form for the holes at the interface between the PbS nanosheet and the metals of higher work function (Au, Pd, and Pt) resulting in p-type behavior. The behavior for gold and palladium contacted devices is well reflected. On the other hand, Schottky barriers arise for the holes at the interface between the PbS nanosheet and the metal of lower work function (Ti). The titanium contact devices still show a slight p-type behavior in contrast to the idealistic view. This might be due to a change in work function of titanium upon oxygen adsorption at the contacts.[46] Also an oxygen adsorption on the PbS nanostructures can lead to an enhanced p-type behavior.[47,48]

Interestingly, the majority of devices contacted with platinum show n-type character (Figure 2d). In contrast to palladium and gold, platinum possess a higher work function. Ohmic contacts for holes will be formed after Fermi level equilibration and Schottky barriers for electrons as displayed in Figure 3. However, compared to the interface PbS/Pd or PbS/Au the formed Schottky barriers for electrons at the interface PbS/Pt are very thin due to the pronounced downwards bending. Positive values for the gate voltage reduce the thickness of the Schottky barriers further. This means that electrons can easily tunnel through the thin Schottky barriers and contribute to the current, resulting in n-type character in the majority of devices contacted with Pt. p-type character in some devices might be due to bad contacts with a wider tunnel barrier hampering the tunneling of electrons. Therefore a hole current is promoted and we measure a slight p-type character.

**Table 1:** *Summary of the key values of the transistor devices: The average conductivity is taken at $V_{DS}$ = +0.25 V and $V_g$ = 0 V and the field-effect mobility at $V_g$ = -5 V (p-type) resp. $V_g$ = +5 V (n-type). The used nanosheets possess an average thickness of 5.4 nm.*

| Contact material | Number of measured devices | Average conductivity [S/m] | Average $I_{ON}/I_{OFF}$ ratio | Average field-effect mobility [cm²/Vs] |
|---|---|---|---|---|
| Ti | 9 | 0.01 | 3.26 | 0.33 |
| Au | 14 | 0.07 | 13.33 | 8.35 |
| Pd | 17 | 0.07 | 9.78 | 7.72 |
| Pt | 27 | 0.10 | 25.29 | 2.57 |

In order to extract photo-generated charge carriers at zero bias (photovoltaic effect) a built-in field is required. In PbS nanosheets contacted by two different metals the built-in field is given by the difference in work functions. Adequate Schottky barriers help to separate the charge carriers. We contacted individual PbS nanosheets with a thickness of 2.4 nm, 5.4 nm, and 11.4 nm via electron-beam lithography in an asymmetric configuration either with palladium and titanium contacts or with platinum and titanium contacts (see AFM images in Figure 1). The purpose of these studies was to investigate the dependence of the photovoltaic performance of the PbS nanosheets on the sheet thickness and on the metal work functions. In order to be able to contract individual nanosheets by e-beam lithography it was necessary to build the devices on Si/SiO$_2$; on glass the precision is not given due to charging. On Si/SiO$_2$ the illumination with white light excited also the direct band gap of the silicon underneath giving incorrect characteristics. Thus, the current-voltage characteristics have been performed under vacuum conditions at room temperature using a 637 nm laser as light source.

In Figure 4 a) and b) the current-voltage characteristics of the PbS nanosheets with 5.4 nm thicknesses are displayed. The curves correspond to dark and illuminated I-V characteristics at different laser output intensities from 2 mW up to 10 mW (10 mW laser intensity corresponds to an illumination power density of 1.17 kW/m²). The I-V curves show a nonlinear behavior with a strong asymmetry and diode-like characteristics become distinct. Upon illumination, the current shifts with increased laser intensities to more negative values. A finite short-circuit current $I_{SC}$ arises at bias voltage $V_{DS}$ = 0 V which indicates photovoltaic effect in the conductive channels of our devices. For an illumination power of 1.17 KW/m² we measure for PbS nanosheet contacted with Pd/Ti a short-circuit current of 134.08 pA in average and an open-circuit voltage of 0.021 V, whereas for PbS nanosheet contacted with Pt/Ti a source-circuit current of 153.42 pA and an open-circuit voltage of 0.066 V are measurable. The photocurrent $I_{SC}$ increases linearly with laser power: A higher photon flux generates linearly more free charge carriers which are harvested by means of the built-in field (Figure 5). On the other hand, the open-circuit voltage $V_{OC}$ tends asymptotically to a saturation level for larger illumination power. The reason is that the maximum $V_{OC}$ that can be reached is proportional to the difference of the work functions of the contact metals and the effective band gap of the semiconductor.[49] Fitting the curve of $V_{OC}$ as a function of laser power $P_{LASER}$ with $V_{OC} = V_{OC,MAX} \cdot (1 - exp(-a \cdot P_{LASER}))$, where $V_{OC,MAX}$ is the asymptotical reached open-circuit voltage and $a$ is a stretching factor, yields $V_{OC,MAX}$ = 0.06 V for Pd/Ti contacted sheets and 0.09 V for Pt/Ti contacted ones. This shows that a larger difference in work function leads to larger maximal reachable open-circuit voltages. The reason that not the full difference in work function or the band gap is extracted as $V_{OC}$ lies in the fact that serial resistances are present. This could be the sheet resistance; but most likely tunnel barriers at the contacts are the main contributors.

From the I-V characteristics we can determine the power conversion efficiency $\eta$ of the PbS nanosheets:

$$\eta = P_{max} / P_{in} = FF \cdot (I_{SC} \cdot V_{OC} / P_{in})$$

, where $P_{max}$ is the maximum extractable electrical power, $P_{in}$ is the incident laser power, and $FF$ is the fill factor.

For PbS nanosheet contacted with Pd/Ti the fill factor $FF$ is found to be 23.26 % averaged over all measured devices and the efficiency $\eta$ lies at 0.04 %. Whereas for PbS nanosheet contacted with Pt/Ti we calculate a fill factor of 32.18 % and an efficiency of 0.23 %. One device shows a maximum efficiency of 0.94 %. Table 2 gives an overview for the key values for the investigated devices. For each metal configuration we investigated at least fifteen devices.

Another key parameter to evaluate the performance is the external quantum efficiency (EQE) which gives the amount of incident photons converted to extracted charge carriers. The EQE can be expressed by the formula:

$$EQE = ( h \cdot c / e \cdot \lambda ) \cdot I_{SC} / P_{in}$$

, where $h$ is the Planck constant, $c$ the speed of light, and $e$ the elementary charge. $P_{in}$ is the illumination power of the explored PbS nanosheet area between the source and drain electrodes.

The average EQE reached in devices contacted with Pd/Ti is 11.60 % whereas PbS nanosheets contacted with Pt/Ti show an averaged EQE of 20.75 %. This can be explained by the fact that with increasing built-in field the photo-generated carriers are more efficiently separated and contribute to the current rather than to recombine.

**Table 2:** *Summary of the key values of the photovoltaic devices.*

| PbS nanosheet thickness | Metal configuration | Number of measured devices | $V_{OC}$ [V] | $I_{SC}$ [nA] | FF | η [%] | EQE [%] |
|---|---|---|---|---|---|---|---|
| 5.4 nm | Pd/Ti | 15 | 0.021 | 0.134 | 0.23 | 0.04 | 11.60 |
| 5.4 nm | Pt/Ti | 32 | 0.065 | 0.153 | 0.32 | 0.23 | 20.75 |
| 2.4 nm | Pd/Ti | 29 | 0.136 | 0.0737 | 0.29 | 0.14 | 10.12 |
| 2.4 nm | Pt/Ti | 12 | 0.189 | 0.0183 | 0.30 | 0.09 | 3.10 |

This does not take into account that not all photons are absorbed by the semiconductor material. For new materials with tunable band gap the absorption is not tabulated. Thus, we estimate the absorption by measuring the transmission *T* in a confocal microscope using a 633 nm laser. This yields an upper limit of the absorption *A = (1-T)* disregarding scattering and reflection which are difficult to access. For 5.4 nm PbS nanosheets we measure an absorption of 43 %. Considering a second absorption after reflection at the substrate silicon (67.51%) yields an IQE of 17.18 % for Pd/Ti contacts and 30.74 % for Pt/Ti ones.

Due to the quantum confinement, the band gap of a 2.4 nm nanosheets is larger than the one with a height of 5.4 nm. By using the particle-in-a-box approach we calculate an effective band gap of 1.54 eV for 2.4 nm nanosheets. In order to study the effects of the size-dependent band gap on the photovoltaic performance we repeat the measurements with the 2.4 nm PbS nanosheets. We observe again a shift in photocurrent at $V_{DS}$ = 0 V with increased laser intensity: For an illumination power density of 1.17 kW/m², the current is 73.72 pA for Pd/Ti contacted and 18.92 pA for Pt/Ti ones. The open-circuit voltage $V_{OC}$ is in average 0.14 V in the case of Pd/Ti-configuration and 0.19 V when using the contact configuration Pt/Ti. Compared to the 5.4 nm thick PbS nanosheets we observe higher open-circuit voltages $V_{OC}$ (also the fitted maximum open-circuit voltages are higher). We studied at least 12 devices for each configuration and we calculate for the Pd/Ti contact configuration an averaged efficiency of 0.14 % with a fill factor of 28.81 % and for the Pt/Ti contact configuration an averaged efficiency of 0.09 % with a FF = 30.44 % as presented in Table 2. EQE amounts to 10.12 % and 3.10 % respectively. With an estimated absorption of 10 % IQE is 53.26 % and 16.32 %, respectively.

Complementary, we measured the I-V characteristics of 11.4 nm nanosheets. They show open-circuit voltages of less than 1 mV and short-circuit currents below 0.1 nA resuling in negligible performances. This is due to an effective band gap which is almost bulk-like.

Exemplarily, we can draw the energy levels schematically for 5.4 nm thick PbS nanosheets in asymmetric configuration with ϕ(Pd) > ϕ(PbS) > ϕ(Ti). Due to the accumulation of positive charges in the Pd/PbS contact region the conduction and the valence band bend downwards. On the other side due to the accumulation of negative charges in the Ti/PbS contact region the conduction and valence band bend upwards. Figure 6 shows the band scheme before and after Fermi level equilibration. According to the band structure scheme photo-generated electrons migrate to the titanium contact and photo-generated holes to the palladium resp. platinum one (without applying a source-drain voltage).

From the data in Table 2 it can be seen that the open-circuit voltage is a function of the nanosheets height, and thus the effective band gap. Large band gaps give rise to higher open-circuit voltages. So do larger differences in the work function of the two contact metals. A higher build-in field leads to a stronger bending of the conduction and the valence band which makes the extraction of charge carriers more efficient. Both contribute to higher power conversion efficiencies. On the other hand, thinner nanosheets have a lower absorption and thus, the short-cut current is lower. The fill factor is not strongly affected by the effective band gap or the contact metal work function. Anyhow, it can be seen that larger differences in the work function of the contact materials increase the fill factor. This is again due to a more efficient separation of the excitons followed by their extraction to the leads resp. a reduced recombination. The power conversion efficiency is a matter of open-circuit voltage, short-cut current, and fill factor. Since those values possess different dependencies on the nanosheets thickness there is an optimum. Though thinner nanosheets yield higher open-circuit voltages, the gain in current in thicker sheets leads to higher efficiencies. Thus, 5.4 nm nanosheets with Pt/Ti contacts give the highest average value of 0.23 %.

We investigated the transistor behavior of individual pristine PbS nanosheets contacted with a variety of metals. Our studies reveal efficient switching of the current by using metal electrodes of higher work function. With palladium as contact material we reach field-effect mobilities of more than 30 cm²/Vs. Furthermore, we observe photovoltaic effect in individually, asymmetrical contacted PbS nanosheets. Our synthesis to control the PbS nanosheets' height allows to investigate the photovoltaic response systematically for two different band gap energies. Together with a high difference in work function of the used contact metals power conversion efficiencies of up to 0.94 % (EQE = 75.70 %) are reached. Though, this is not yet at the level of similar devices based on carbon nanotubes[50,51] with their extraordinary mobilities, PbS nanosheets outperform CdSe ribbons.[52] This shows that solution processable PbS nanosheets are promising as energy converter material for future low-cost solar cells.

**Experimental Section**

**Syntheses.** All chemicals were used as received but the 1,1,2-trichloroethane (TCE) which was distilled to remove the isopropanol which is added for stabilization by the manufacturer. The chemicals used were lead(II) acetate tri-hydrate (Aldrich, 99.999%), thioacetamide (Sigma-Aldrich, >= 99.0%), diphenyl ether (Aldrich, 99%+), dimethyl formamide (Sigma-Aldrich, 99.8% anhydrous), oleic acid (Aldrich, 90%), 1,1,2-trichloroethane (Aldrich, 96%), 1,1-dichloro-3,3-dimethylbutane (Aldrich 99%).

*Lead oleate precursor:* A three neck 50 mL flask was used with a condenser, septum and thermocouple. 860 mg of lead acetate trihydrate (2.3 mmol) were dissolved in 10 mL of diphenyl ether and 2 mL or 3.5 mL of OA (5.7 mmol; 9.9 mmol) and heated to 75 °C until the solution turned clear. Then a vacuum was applied for 3.5 h to transform the lead acetate into lead oleate and to remove the acetic acid in the same step.

*Lead sulfide nanosheets (2.4 nm thickness):* The lead oleate solution was heated under nitrogen flow rate to a reaction temperature of 130 °C while at 100 °C 1 mL of TCE was added under reflux to the solution and the time has been started. After 12 minutes 0.23 mL of a 0.04 g TAA (0.5 mmol) in 6.5 mL DMF were added to the reaction solution. After 5 minutes the heat source was removed and the solution was let to cool down. Afterwards the solution was centrifuged at 4000 rpm for 3 minutes. The precipitant was washed two times in toluene before the nanosheets were finally suspended in toluene for storage.

*Lead sulfide nanosheets (5.4 nm thickness):* The lead oleate solution was heated under nitrogen flow rate to a reaction temperature of 160 °C while at 100 °C 1,25 mL of 1,1-dichloro-3,3-dimethylbutane was added under reflux to the solution and the time has been started. After 12 minutes 0.2 mL of a 0.04 g TAA (0.5 mmol) in 6.5 mL DMF were added to the reaction solution. After 5 minutes the heat source was removed and the solution was let to cool down. Afterwards the solution was centrifuged at 4000 rpm for 3 minutes. The precipitant was washed two times in toluene before the nanosheets were finally suspended in toluene for storage.

**Band Level Calculations.** To calculate effective band gap of the PbS nanosheets we used the particle-in-a-box approach with the effective mass approximation. In x and y direction the nanosheets are not in confinement, only in z direction. The band levels are calculated using the expression $E_{cond,eff} = E_{cond,bulk} - h^2/8m_eL_z^2$ for the conduction band level with the effective conduction band level $E_{cond,eff}$, the bulk conduction band level $E_{cond,bulk}$ = 4.6 eV,[46] the Planck constant $h$, the effective mass for electrons in the material $m_e^* = 0.12\ m_e$, and the height of the nanosheets $L_z$. For holes a corresponding formula was used ($E_{val,bulk}$ = 5.0 eV, $m_h^* = 0.11\ m_e$). The effective band gap is calculated using the expression $E_{eff} = E_{g,bulk} + (h^2/8m_{red}L_z^2)$, where $m_{red} = (1/m_e^* + 1/m_h^*)$. The work functions of the metals are taken from: Karl W. Böer – Survey of semiconductor physics, Vol. II.

**Device Preparations.** As substrate we used silicon wafers covered with 300 nm thermal silicon oxide as gate dielectric. The highly doped silicon was used as backgate. PbS nanosheets suspended in toluene were then spin-coated on those samples and individual nanosheets were contacted by e-beam lithography followed by thermal evaporation of the metals titanium, gold, palladium and lift-off. In the case of the titanium contacts to avoid the oxidation we deposit an additional layer of a few nanometers of gold on top of the titanium.

**Electrical Characterization.** Immediately after device fabrication we transferred the samples to a probe station (Lakeshore-Desert) connected to a semiconductor parameter analyzer (Keithley 4200). The transfer and output characteristics have been performed in vacuum at room temperature.

The measurements were performed right after the transfer to vacuum. In order to check the stability of the devices some samples were measured again after three weeks. They showed a current drop of less than one order of magnitude. After that no changes were observed anymore.

The current-voltage characteristics of the photovoltaic devices have been performed with and without illumination using a 637 nm laser. The spot size of the laser was 3 mm in diameter. Thus, all devices were fully illuminated since the spacing between source and drain electrodes was in average around 600 nm. The illuminated area between the source and drain electrodes was determined using AFM images which were also considered for the calculations of performance values. For the estimation of the absorption a confocal microscope Olympus IX81 with a laser wavelength of 633 nm was used.

**Transmission Electron Microscopy.** TEM characterization was performed on a JEOL-1011 microscope with an acceleration voltage of 100 kV. The TEM samples were prepared by diluting the nanosheet suspension with toluene and then drop casting 10 µL of the suspension on a TEM copper grid coated with a thin carbon film.

**Atomic Force Microscopy.** The measurements were performed on a Veeco Dimension 3000 AFM in contact mode. The samples were prepared by spin-coating the nanosheet suspension on a silicon wafer.

**Acknowledgements**

The authors thank the German Research Foundation DFG for financial support in the frame of the Cluster of Excellence „Center of ultrafast imaging CUI". For the funding from the European Research Council (Seventh Framework Program FP7, Project: 304980 / ERC Starting Grant 2D-SYNETRA) the authors are also grateful. CK acknowledges the German Research Foundation DFG for a Heisenberg scholarship (KL 1453/9-1). Furthermore, the authors thank Alexander Gräfe for helpful discussions.

**FIGURES**

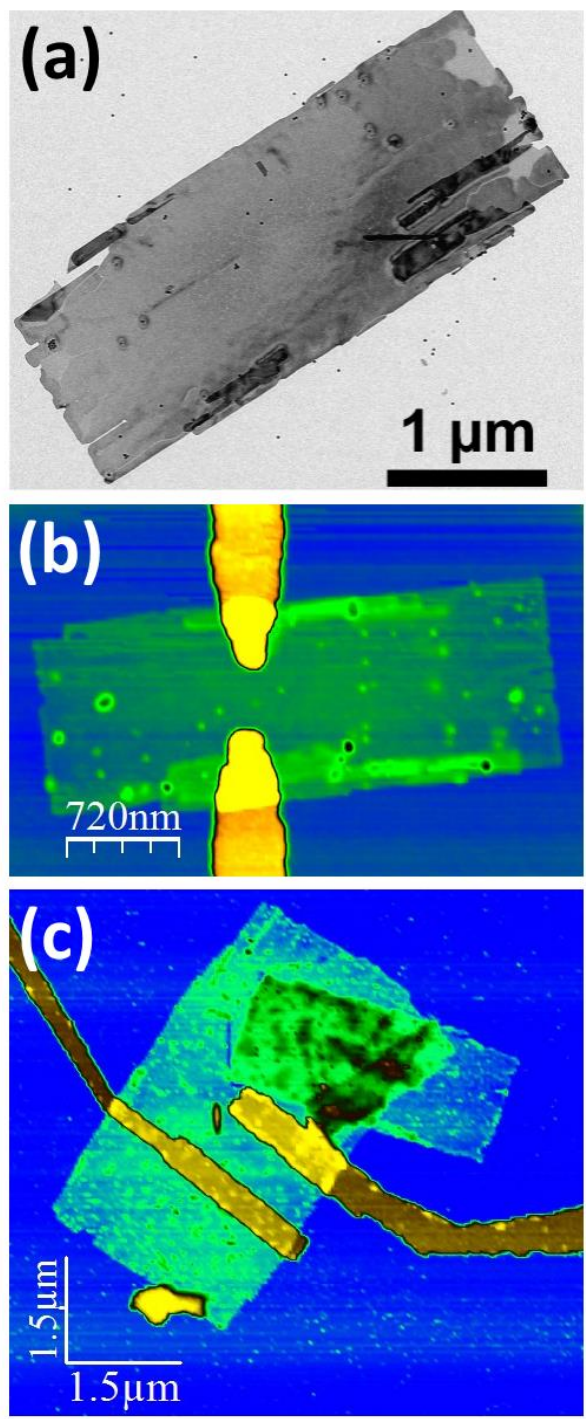

**Figure 1.** *(a) TEM image showing an individual PbS nanosheet. (b) AFM image of an individual PbS nanosheet contacted with Au leads. The PbS nanosheet possesses a total height of 9 nm. This includes a top and bottom layer of self-assembled oleic acid with a thickness of about 1.8 nm, such that the inorganic part has a height of 5.4 nm. (c) AFM image of an individual PbS nanosheet contacted with a palladium and a titanium electrodes for photovoltaic measurements.*

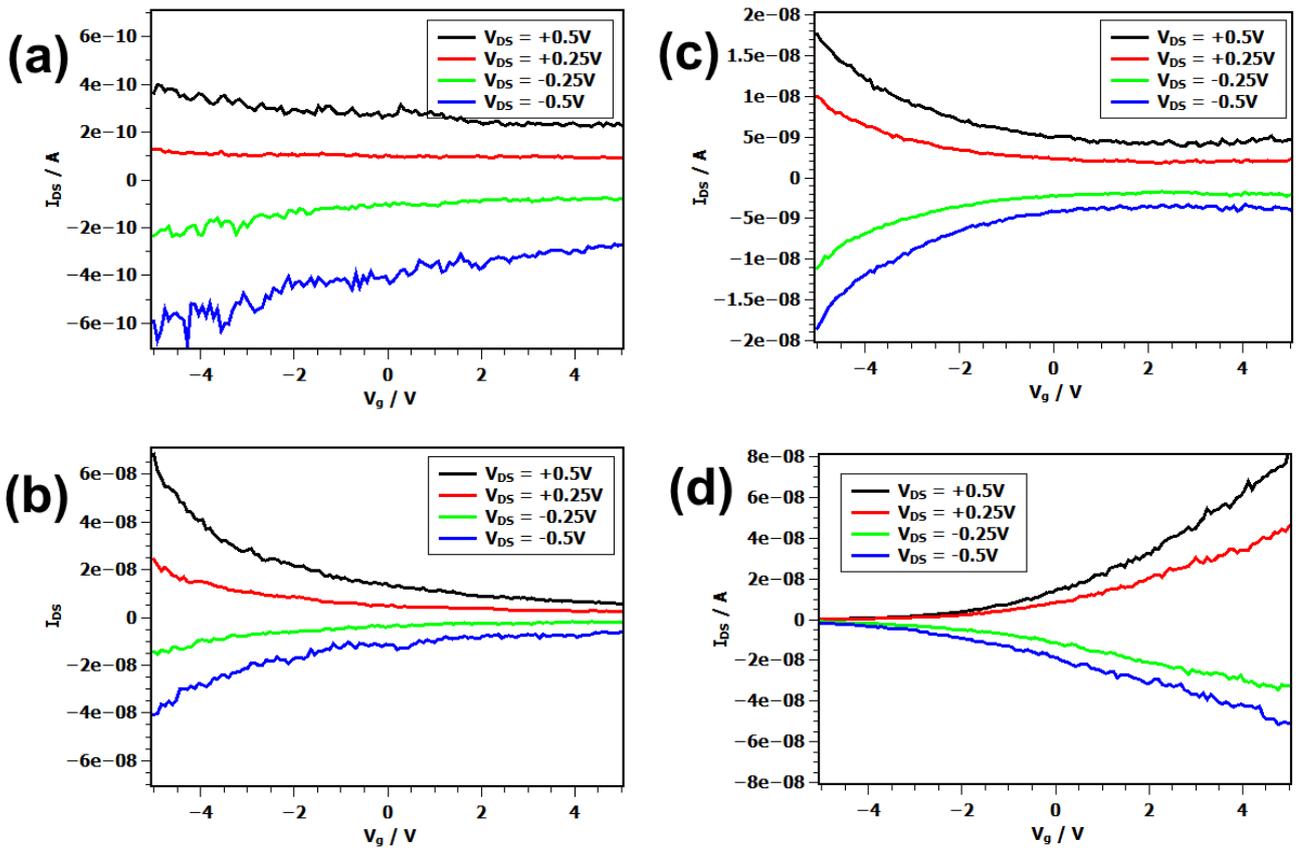

**Figure 2.** *Transfer characteristics under various bias voltages of 5.4 nm PbS nanosheet transistor using (a) titanium, (b) gold, (c) palladium, and (d) platinum electrodes.*

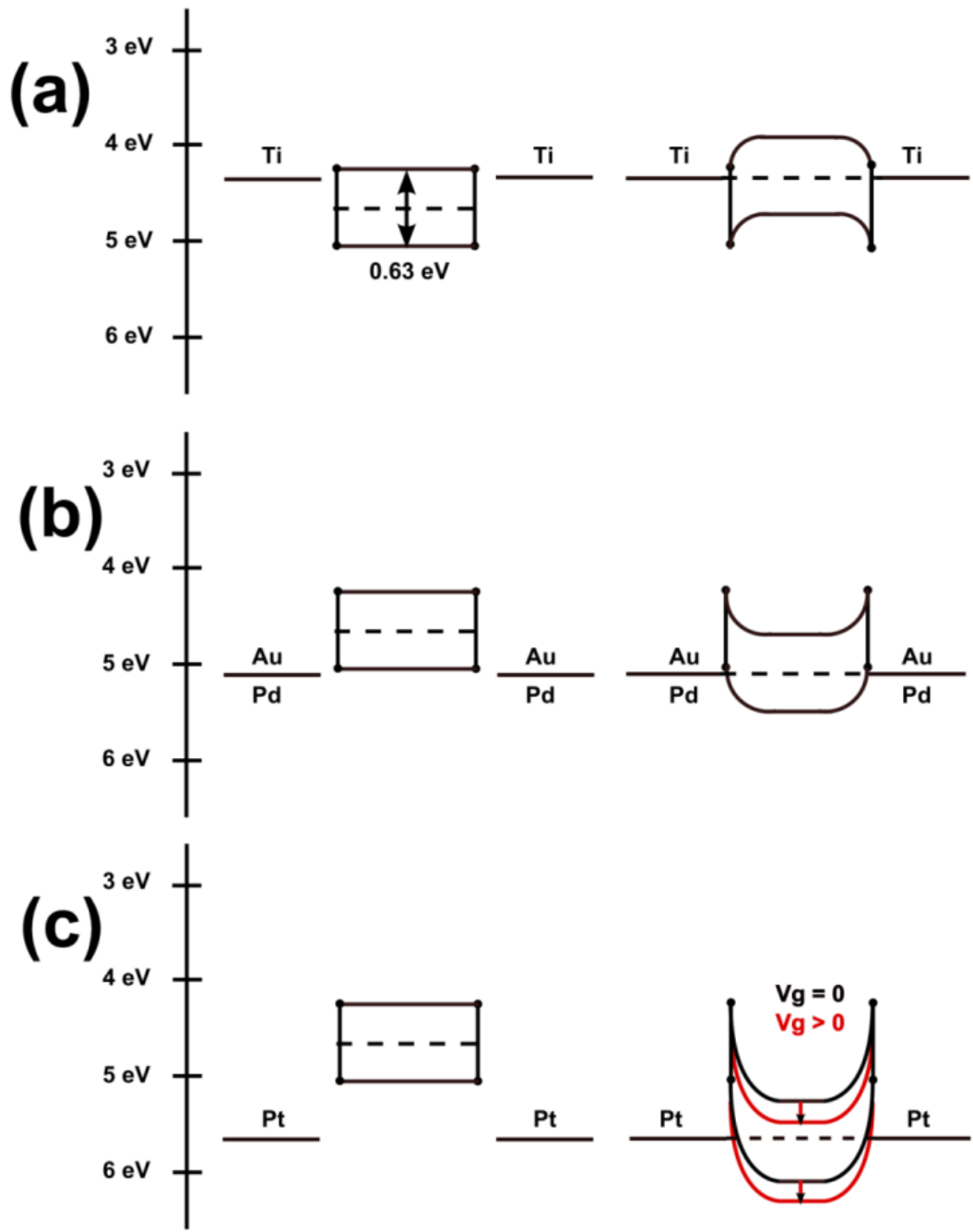

**Figure 3.** *Schematic band alignment of 5.4 nm thick PbS nanosheets contacted in a symmetrically configuration with (a) titanium, (b) gold or palladium, and (c) platinum before and after Fermi level equilibration.*

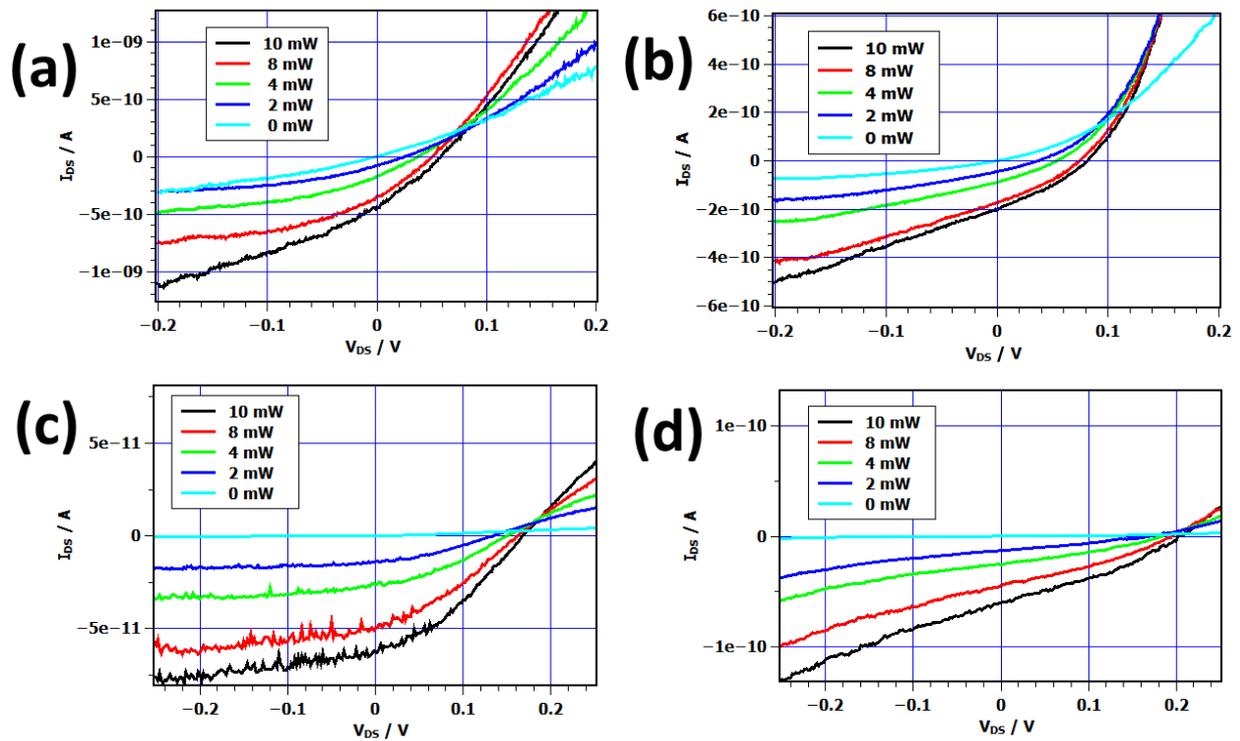

**Figure 4.** *I-V characteristics performed in the dark and under illumination for various intensities for a) a 5.4 nm thick PbS nanosheet contacted with palladium and titanium, b) with platinum and titanium, c) a 2.4 nm thick PbS nanosheet contacted with palladium and titanium, d) with platinum and titanium ($V_g$ = 0 V).*

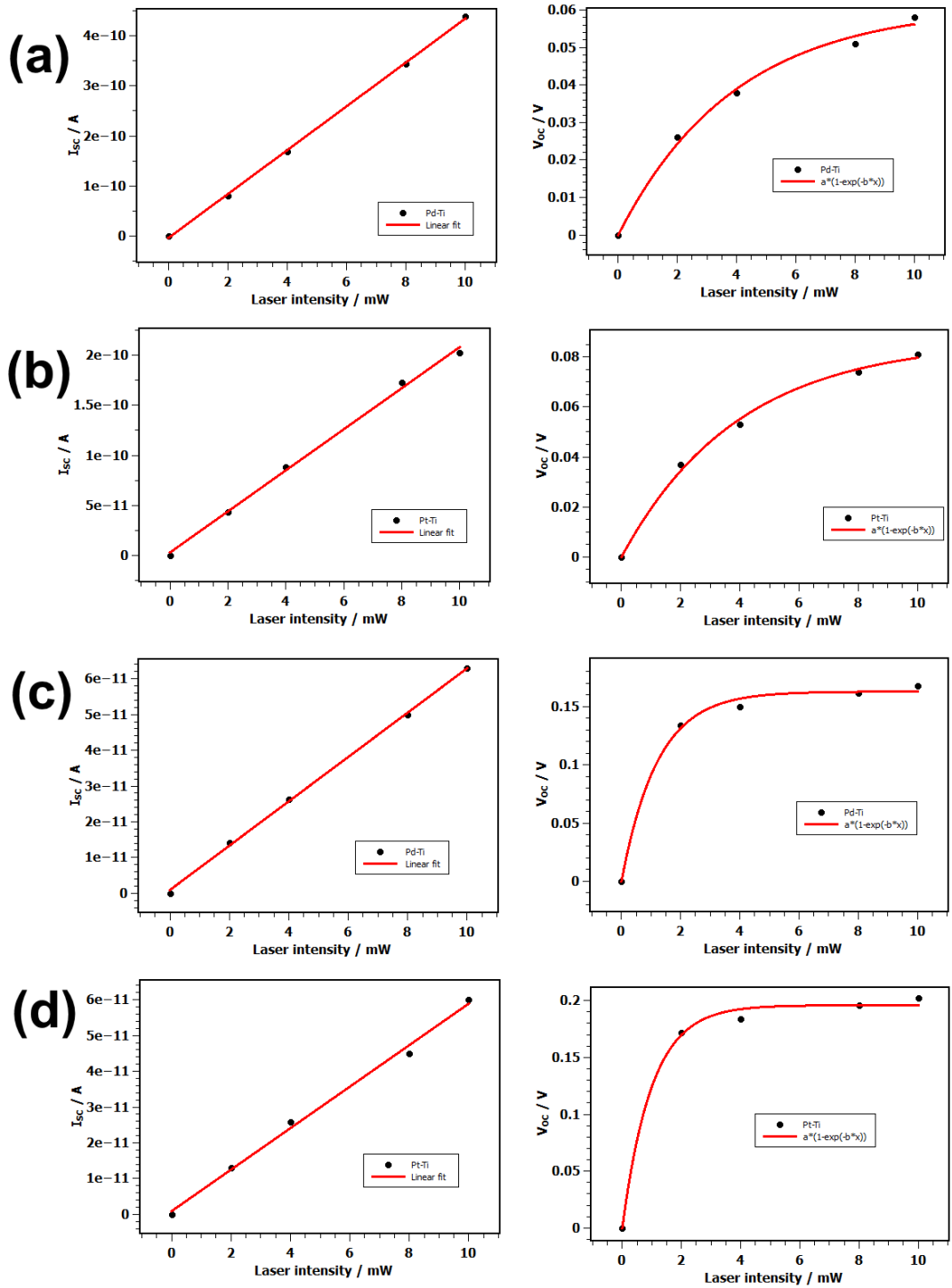

**Figure 5.** *Shortcut current (left side) and open-circuit voltage (right side) as a function of the laser intensity $P_{LASER}$. The currents are fitted linearly and the voltages with a function $V_{OC} = V_{OC,MAX} \times (1 - \exp(-a \times P_{LASER}))$, where $V_{OC,MAX}$ is the asymptotical reached open-circuit voltage and a is a stretching factor. a) 5.4 nm thick PbS nanosheet contacted with palladium and titanium ($V_{OC,MAX}$ = 0.06 V) and b) contacted with platinum and titanium ($V_{OC,MAX}$ = 0.09 V). c) 2.4 nm thick PbS nanosheet contacted with palladium and titanium ($V_{OC,MAX}$ = 0.16 V) and d) with platinum and titanium ($V_{OC,MAX}$ = 0.20 V).*

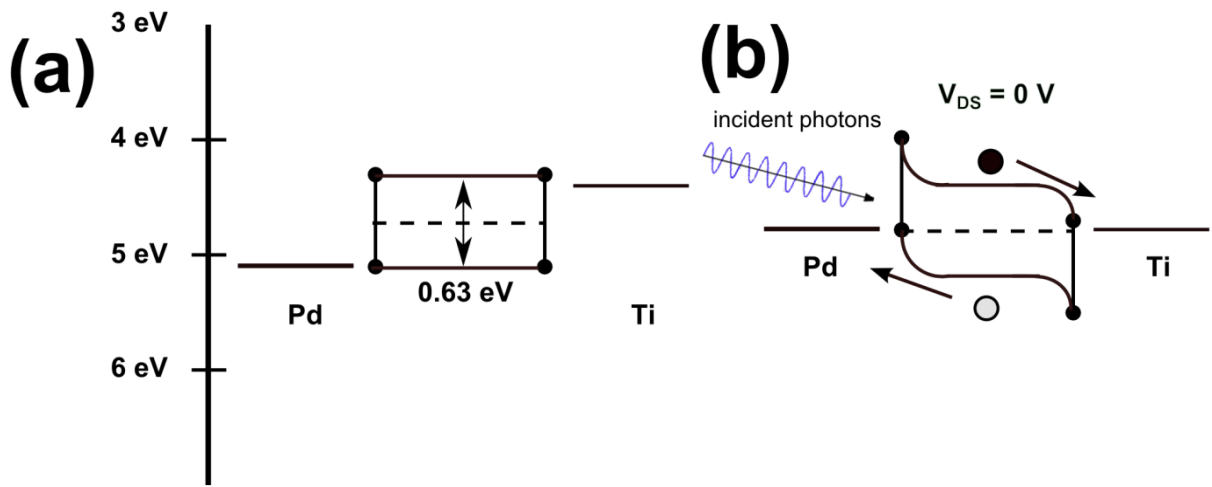

**Figure 6.** *Schematic band alignment of 5.4 nm thick PbS nanosheets contacted with palladium and titanium (a) before and (b) after Fermi level equilibration (black dot – electron, grey dot – hole).*

**Figure Table-of-contents**

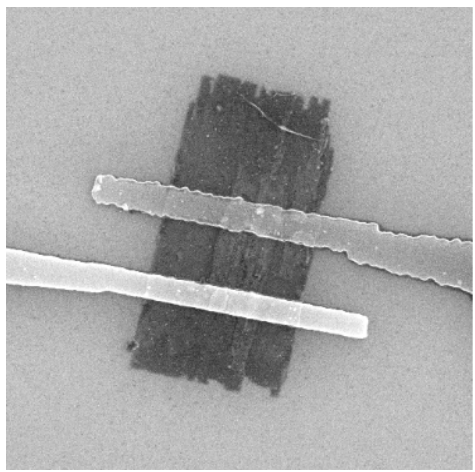 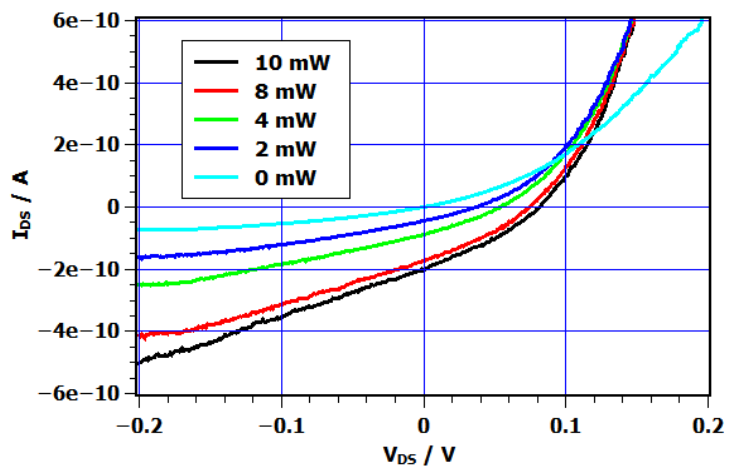

*Asymmetrically contacted lead sulfide nanosheets show photovoltaic effect. The nanosheets' height has a decisive impact on the device performance.*